# Against 'Reality' in Physics[*]

Stephen Boughn[†]

Department of Physics, Princeton University, Princeton NJ
Departments of Physics and Astronomy, Haverford College, Haverford PA

**Abstract**

The concept of *reality* is often raised in the context of philosophical foundations of physics or interpretations of quantum mechanics. When this term is so raised, it is a warning to me that I'm about to be led down a rabbit hole. Such diversions usually lead nowhere unless you consider endless discussions of Schrödinger's cat, wave function collapse, quantum non-locality, and parallel universes to be useful. A prime example is the famous Einstein, Podolsky, and Rosen paper wherein they concluded that the quantum wave function cannot provide a complete description of physical *reality*. In this essay, I suggest that, in physics discourse, the term 'reality' should be avoided at all costs.

**Against Measurement**

Let me begin by reminding you of a paper by John S. Bell. No, not that one. I'm not talking about his 1964 paper "On the Einstein Podolsky Rosen (EPR) Paradox", in which he introduced his now famous inequality; although, we'll get to that paper as well as the EPR paper momentarily. I'm referring to Bell's 1990 paper entitled "Against 'Measurement'". In this paper, Bell lamented how "uncertainty over terms such as 'apparatus' is still rife in serious discussions of quantum mechanics". Bell sought an "exact formulation" of quantum mechanics "with nothing left to the discretion of the theoretical physicist". He maintained that some words have no place in such a formulation, words such as *system, apparatus, environment, microscopic, macroscopic, reversible, irreversible, observable, information,* and *measurement*. According to Bell, the worst of these is "measurement", hence the title of his paper;

The first charge against 'measurement', in the fundamental axioms of

---

[*] This essay should be considered a companion to "What are Physical States?" (Boughn 2019)
[†] sboughn@haverford.edu

> quantum mechanics, is that it anchors there the shifty split of the world
> into 'system' and 'apparatus'…The word has had such a damaging effect
> on the discussion, that I think it should now be banned altogether in
> quantum mechanics.

The shifty split to which Bell is referring is the Copenhagen interpretation's maxim that while microscopic systems are described by quantum mechanics, macroscopic experiments are to be described by classical physics. Even though Bell offended my experimentalist's sensibilities with such phrases as "piddling laboratory operations" and "the petty world of the laboratory", I agree with him that careless use of some of the above terms has contributed to the conundrums encountered when grappling with the foundations of quantum mechanics.

Bell, clever man that he was, did not directly refer to the "reality" of the natural world, the term that I suggest should be avoided at all costs; however, he did, in effect, put this word into the mouths of others.[1] Still, one get's the feeling that this concept is lurking just below the surface in his paper. When he remarks,

> But experiment is a tool. The aim remains: to understand the world. To
> restrict quantum mechanics to be exclusively about piddling laboratory
> operations is to betray the great enterprise. A serious formulation will
> not exclude the big world outside the laboratory,

I have the impression he means that quantum mechanics should not be simply a theory that predicts the results of measurements, but rather, a theory that provides a description of reality. One might likewise accuse me of putting words in Bell's mouth and that might, indeed, be the case. However, in the very first paragraph of his paper he opines,

> Surely, after 62 years, we should have an exact formulation of some
> serious part of quantum mechanics…I mean too, by 'serious', that
> 'apparatus' should not be separated off from the rest of the world into
> black boxes, as if it were not made of atoms and not ruled by quantum
> mechanics.

He clearly thinks that the current formulation of quantum mechanics is lacking.

The bulk of his paper is then devoted to "the measurement problem" and the implied "collapse of the wave function", dilemmas that arise in the context of ascribing reality to quantum states. He proceeds with discussions of commentaries by Dirac,

---

[1] For example, he declares that Gottfried supposes that wave function collapse "really" takes place and that van Kampen identifies "realities" that characterize the results of experiments.

Landau, Gottfried, von Neumann, and van Kampen, critiques with which I heartily agree. However, whereas Marcel Reginatto and I concluded that these dilemmas are faux problems (Boughn & Reginatto 2013), Bell clearly thinks they are legitimate questions to be answered. This is made abundantly clear in the final section of the paper entitled, "Towards a precise quantum mechanics", in which he declares,

> It seems to me that the only hope of precision with the dual ($\Psi, x$) kinematics is to omit completely the shifty split, and let both $\Psi$ and $x$ refer to the [real] world as a whole, where $\Psi$ denotes a quantum state and $x$ classical variables (and where I have, again, put a word in Bell's mouth). This is precisely the picture presented in de Broglie-Bohm mechanics. Bell also points out another way to avoid the vagueness of wave function collapse and that is to modify Schrödinger's equation to include a stochastic collapse mechanism as in the Ghiradi-Rimini-Weber (GRW) scheme. In the end Bell concludes that currently (as of 1990):

> At least two roads are open from there [orthodox theories] towards a precise theory, it seems to me. Both eliminate the shifty split [between quantum system and measurement apparatus]. The de Broglie-Bohm type theories retain, exactly, the linear wave equation, and so necessarily add complementary variables to express the non-waviness of the world on the macroscopic scale. The GRW-type theories have nothing in their kinematics but the wavefunction. It gives the density (in a multidimensional configuration space!) of stuff. To account for the narrowness of that stuff in macroscopic dimensions, the linear Schrödinger equation has to be modified, in the GRW picture by a mathematically prescribed spontaneous collapse mechanism.

I maintain that Bell has, indeed, led us down a rabbit hole, that is, unless you consider de Broglie-Bohm mechanics or the GRW statistical collapse theory to be eminently reasonable approaches.

**EPR Paradox**

The 1935 Einstein, Podolsky, and Rosen paper, "Can Quantum-Mechanical Description of Physical Reality Be Considered Complete?", was notable in that it gave an explicit criterion of *reality*. I first read the EPR paper when I was in graduate school and was mystified. I had assumed the paper would present an argument about the self-consistency of quantum mechanics or lack there of. Instead, it seemed to be less about physics and more about the philosophical issue of what constitutes reality. The conclusion of EPR was quite simply that quantum mechanics "does not provide a complete description of the physical reality." In fact, the paper was penned by Podolsky and Einstein was not happy with it (Howard 2007).

Howard points out that in a letter to Schrödinger, written a month after the EPR paper was published, Einstein chose to base his argument for incompleteness on what he termed the "separation principle" and continued to present this argument "in virtually all subsequent published and unpublished discussions of the problem". According to the separation principle, the *real state of affairs* in one part of space cannot be affected instantaneously or superluminally by events in a distant part of space. Suppose *AB* is the joint state of two systems, *A* and *B,* that interact and subsequently move away from each other to different locations. (Schrödinger (1935) would later introduce the term *entangled* to describe such a joint state.) In his letter to Schrödinger, Einstein explained (Howard 2007)

> After the collision, the real state of (*AB*) consists precisely of the real state *A* and the real state of *B*, which two states have nothing to do with one another. *The real state of B thus cannot depend upon the kind of measurement I carry out on A* [separation principle]. But then for the same state of *B* there are two (in general arbitrarily many) equally justified [wave functions] $\Psi_B$, which contradicts the hypothesis of a one-to-one or complete description of the real states.

His conclusion was that quantum mechanics does not provide a complete description of reality. Note that Einstein's separation principle did not claim that a measurement of system *A* has no effect on the *result of any measurement* on system *B* but rather that a measurement of system *A* has no effect on the *real state* of system *B*. So even in his revised argument, Einstein relies on the notion of the "real state" of a system, that is, the notion that the state of a system constitutes an aspect of reality. Again, we've been led down a rabbit hole. If Einstein's separation principle had stopped at "a measurement of system *A* has no effect on the *result of any measurement* on system *B*", a principle that is completely consistent with standard quantum theory, then there would be no basis for his argument that quantum mechanics does not provide a complete description of reality.[2]

**The Many-Worlds Interpretation**

There is another theory, Hugh Everett's so-called *many worlds interpretation*, that directly addresses both the shifty split between the quantum system and measurement apparatus and the collapse of the wave function.[3] This interpretation is considered by some to be mainstream (but certainly not by me). Everett sought to remove the shifty split as well as eliminate wave function collapse by postulating a universal wave function that never collapses but continues to evolve according to the Schrödinger equation. Furthermore, the universal wave function describes everything in the universe, including the measuring apparatus and observer of that apparatus. The standard interpretation, introduced by DeWitt and Graham (1973), is that whenever a measurement is made all possible outcomes are realized, each in a different universe, i.e., the universe splits into many. The universal wave function describes the entire multiverse of these many worlds. I agree with van Kampen (see footnote 3) that this is a "mind-boggling fantasy" and certainly qualifies as descending down a rabbit hole. On the other hand, Everett did not characterize his "theory of the universal wave function" as implying the existence of a multiverse consisting of many real worlds. In fact, Everett's theory is much more nuanced and it is telling that in his theory of the universal wave function, the words "real" and "reality" only appear within quotation marks (DeWitt & Graham 1973). He seems to be warning us about the danger of taking these terms too literally. In an appendix he even advises us to avoid the tendency of identifying highly successful physical models with "reality" itself.[4] So I will refrain from accusing Everett of leading us down a rabbit hole; however, the subsequent expounders of his theory certainly have.

**Bell's Theorem and Quantum Non-locality**

Bell's 1964 paper on the EPR paradox is considered by some to be one of the most important papers ever written on the foundations of quantum mechanics. My assessment is more modest as I have expressed elsewhere (Boughn 2016, 2017). Bell characterized the EPR argument as: "an argument that quantum mechanics could not be a complete theory but should be supplemented by additional variables. These additional variables were to restore to the theory causality and locality." He then proceeds to add hidden variables that restore causality and locality but demonstrates that the predictions of the resulting theory satisfy an inequality (Bell' theorem) that is inconsistent with the predictions of quantum mechanics. While Bell's is a mischaracterization of EPR's claim[5], it certainly captures what many felt was a serious problem with quantum mechanics. In fact, by 1927 Einstein had already attempted a hidden variable model of Schrödinger's wave mechanics and failed because he "could not find a way around the entanglement" (Howard 2007). I suspect Einstein would not have been very impressed by Bell's proof. Like his 1990 paper, Bell's 1964 paper never mentions "reality", except for a quote from EPR. However, again the notion of reality lurks just below the surface. For what are these hidden variables but properties of an underlying physical reality? They have well defined classical values not subject to the vagaries of quantum mechanics.

---

[2] On the other hand, I quite agree that quantum mechanics does not provide a complete description but for entirely different reasons (Boughn 2017, Boughn & Reginatto 2019).

[3] Bell didn't comment on this theory directly in his 1990 paper but quotes van Kampen's assessment noting that van Kampen has no patience for "such mind-boggling fantasies as the many world interpretation".

[4] I've discussed Everett's theory in detail elsewhere. (Boughn 2018)

The only way Bell could mitigate his conclusion was by introducing "…a mechanism whereby the setting of one measuring device can influence the reading of another instrument, however remote. Moreover, the signal involved must propagate instantaneously, so that such a theory could not be Lorentz invariant." Of course, Einstein would have surely rejected this possibility[6]. Even so, this was not the end of Bell's quantum quandaries. Recall, that he was extremely bothered by the "shifty split" between quantum system and classical measuring apparatus and by the dilemma of wave function collapse.

He sought to resolve both of these with his "theory of local beables" (Bell 1975).

---

[5] Rather, EPR claimed that "…the wave function does not provide a complete description of the reality…" but they "…left open whether or not such a description exists. [They] believe, however, that such a theory is possible."

[6] Although, it should be noted that de Broglie-Bohm mechanics, which Bell was intrigued by, did predict such superluminal signals.

Bell endeavored to precisely embed classical notions in quantum mechanics in a way that observables cannot be. "Beables" provided such a device. The very name "beable" indicates that Bell fancies them as part of an underlying physical reality; although, of course, he doesn't use the word "reality". He comes close when he designates beables as "physical" and wave functions as "non-physical". In an analogy with electromagnetic fields, which are physical, and electromagnetic potentials, which are non-physical, Bell indicates that it doesn't matter that the electromagnetic scalar potential propagates with infinite velocity because "It is not *really* supposed to be there. It is just a mathematical convenience." [my italics] Again, the notion of *reality* is just below the surface. These beables take the place of hidden variables in his 1964 paper and so it's not surprising that he concludes that this (potentially) precise formulation of quantum mechanics demonstrates that there is a "gross non-locality of nature".[7] So again, attributing "reality" to some aspect of nature has led us down a rabbit hole. In the next section, I'll elaborate on just why this is the case.

**Why 'Reality' Doesn't Belong in Physics**

You will have noticed that of the 5 works I referred to above, only one, the EPR paper, referred to the *reality* of the physical world. This was partly by design. I wanted to draw your attention to the fact that even when "reality" isn't directly invoked, it's often lurking just below the surface. On the other hand, in more philosophical works about the foundations of quantum mechanics, physical reality is often right up front. For example, in the collection of essays by philosopher Alyssa Ney and physicist-philosopher David Albert, *The Wave Function: Essays on the Metaphysics of Quantum Mechanics* (2013), 6 of the 10 essays have either *realism* or *ontology* in their titles and their common theme concerns the reality of the Schrödinger wave function. In the popular press, the nature of physical reality is a recurrent theme. One of my friends recently sent me a *PBS Digital Studios* YouTube video (1.5 million hits), "How the Quantum Eraser Rewrites the Past". The first words out of the narrator's mouth were "Can *reality* be adjusted after events have occurred?" The outcomes of these forays invariably lead to trips down a rabbit hole.

So what's my resolution of the problems of wave function collapse and the shifty split between a quantum system and a classical measurement apparatus? I maintain these are quandaries that don't need resolution. Physical theories are models, human inventions, that help us to understand nature as well as to build new devices and make predictions based on these models. In quantum mechanics, wave functions are used to specify the state of a system. Wave function collapse is simply a mnemonic for updating a wave function to account for new information about a system, "just as it would be in classical statistical theory" (Stapp 1972). The problem of the shifty split between quantum system and classical measurement seems more complicated; however, Bohr didn't seem to be bothered by it. Let's see why. Consider his following brief description of a measurement (Bohr 1963, p. 3):

> The decisive point is to recognize that the description of the experimental arrangement and the recordings of observations must be given in plain language, suitably refined by the usual terminology. This is a simple logical demand, since by the word 'experiment' we can only mean a procedure regarding which we are able to communicate to others what we have done and what we have learnt.

---

[7] Although, Bell proves that this non-locality cannot be used for superluminal communication.

Nowhere in this description does he refer to classical physics. In fact, descriptions of measurements are not included in either quantum or classical physics. They are part of Bohr's "procedure regarding which we are able to communicate to others what we have done and what we have learnt." If there is a shifty split between physical theory and measurement in quantum theory, it's also present between classical theory and measurement. Because physicists have long since become comfortable with the relation between theory and measurement in classical physics, perhaps the quantum case shouldn't be viewed as particularly problematic.[8]

To be sure, probing well-established theories to discover previously unknown physical phenomena is a tried-and-true enterprise in physics; however, these endeavors involve specific observable effects predicted by physical theories. Such predictions are possible because of the well-defined meaning of statements made within the context of the theories. On the other hand, the invention of new theories are usually motivated by new experiments and observations that either contradict or fail to be described by previous theories. To be sure, there are other legitimate reasons for seeking to create new theories, for example, unification and simplification. The search for a quantum theory of gravity falls in this category; although, as an experimentalist, I have less interest in efforts that don't involve experimental verification.

---

[8] Marcel and I have dealt with this topic in more detail elsewhere (Boughn & Reginatto 2013, 2019). Einstein (1936) was also well aware of this situation and seemed not to be bothered by it in the least.

However, questions involving the "reality" of aspects of a theory are not of this sort. Reality is a perfectly respectable topic in everyday conversation. What's the harm in inquiring about the reality of the natural world? In short, the harm is that these questions invariably lead nowhere. That such quandaries as wave function collapse, the quantum classical divide, Schrödinger's cat, and the non-locality of entangled quantum states have remained unresolved after more than 80 years is evidence of this fact. Furthermore, the resolutions advanced to solve these reality problems invariably serve no pragmatic purpose. Standard quantum mechanics is quite capable of handling all of the scenarios mentioned above, including the quantum eraser. There are no unanswered physics questions that need be addressed in this way.

Am I an idealist , or perhaps, a logical positivist? I would say not. I take for granted that there's a real physical world out there; after all, I experience it every morning when I wake up. It's the job of physicists to try to understand that reality by creating physical models that can be verified through experiment and observation. That's all I need to know about reality. One might argue that pursuing such unresolvable metaphysical issues can lead to the creation of useful new theories. I have no doubt that many physicists' physical intuition comes about by trying to imagine why the natural world is the way it is. They subsequently use this physical intuition to create new theories that can be subsequently tested experimentally. I'm quite prepared to admit that this is sometimes, or even often, the case. On the other hand, an examination of the sources of scientific creativity is certainly beyond my poor powers and I won't attempt to address that topic here.

Note added in proof: After posting this paper, Hrvoje Nikolić drew my attention to a similarly entitled paper, "Against 'Realism'", by Travis Norsen (2007). While this paper was similarly motivated by EPR and Bell's theorem (with a title similarly motivated by Bell's 1990 paper), Norsen was exclusively interested in the misuse of the notion of "local realism" in discussions of Bell's theorem and, unlike my advice, he did "not think the word [realism] should be banned altogether". Norsen also accepts the manifest non-locality of quantum mechanics and considers wave function collapse to be a crucial quantum postulate, both with which I disagree.

**Acknowledgements**

I gratefully acknowledge Freeman Dyson and Marcel Reginatto for their inspiration and willingness to talk to me about such things.